\documentclass[a4paper,11pt,abstract=true]{scrartcl}

\usepackage{libertine}
\usepackage{hyperref}

\usepackage[scaled=0.81]{beramono}
\usepackage{xparse,xspace}

\NewDocumentCommand{\Tool}{m}{\texttt{#1}\xspace}
\newcommand{\klee}{\Tool{KLEE}}
\newcommand{\AFL}{\Tool{AFL}}

\newcommand{\CBMC}{\Tool{CBMC}}
\newcommand{\code}[1]{\mbox{\lstinline{#1}}}

\newcommand{\tr}[1]{\ensuremath{\textsf{#1}}}

\NewDocumentCommand{\hp}{}{HyperPUT\xspace}

\usepackage{fontawesome}
\newcommand{\setIn}{{\faCircle}}
\newcommand{\setOut}{{\color{lightgray}\faCircleO}}

\usepackage{color} 

\definecolor{redcol}{RGB}{215,25,28}
\definecolor{orangecol}{RGB}{253,174,97}
\definecolor{lightbluecol}{RGB}{171,217,233}
\definecolor{darkbluecol}{RGB}{44,123,182}
\usepackage{numprint}

\usepackage[inline]{enumitem}

\usepackage{amsmath}
\usepackage{tikz}
\usetikzlibrary{shapes,arrows,positioning,calc,fit,intersections,through,graphs, bending} 

\usepackage{svg}
\usepackage[linesnumbered,lined,commentsnumbered]{algorithm2e}

\usepackage[T1]{fontenc}
\usepackage{float}

\usepackage{listings}
\usepackage{graphicx} 
\usepackage{booktabs,multirow} 
\usepackage{hyperref} 
\usepackage{xurl}
\usepackage{tabularx}

\usepackage{csvsimple}
\usepackage{longtable}
\usepackage{booktabs}
\usepackage{caption}
\usepackage{subcaption}

\providecommand{\keywords}[1]
{
  \small	
  \textbf{\textit{Keywords---}} #1
}

\title{HyperPUT: \\Generating Synthetic Faulty Programs \\to Challenge Bug-Finding Tools}

\author{Riccardo Felici $\ \cdot\ $ Laura Pozzi  $\ \cdot\ $ Carlo A. Furia\\[2mm]
  {\normalsize USI Università della Svizzera italiana, Lugano, Switzerland}}

\begin{document}

\maketitle

\begin{abstract}
%\abstract{
  As research in automatically detecting bugs grows and produces new
  techniques, having suitable collections of programs with known bugs
  becomes crucial to reliably and meaningfully compare the effectiveness of
  these techniques.
  Most of the existing approaches rely on \emph{benchmarks} collecting 
  manually curated real-world bugs, or synthetic bugs seeded into real-world programs.
  Using real-world programs entails that extending the existing benchmarks
  or creating new ones remains a complex time-consuming task.

  In this paper, we propose a complementary approach that automatically generates
  programs with seeded bugs.
  Our technique, called \hp, builds C~programs from a ``seed'' bug
  by incrementally applying program transformations 
  (introducing programming constructs such as conditionals, loops, etc.)
  until a program of the desired size is generated.
  In our experimental evaluation, we demonstrate how \hp can generate
  buggy programs that can challenge in different ways the capabilities of
  modern bug-finding tools, and some of whose characteristics are comparable
  to those of bugs in existing benchmarks.
  These results suggest that \hp can be a useful tool
  to support further research in bug-finding techniques---in particular their
  empirical evaluation. 
 % }
\end{abstract}


\section{Introduction}

Research in 
detecting bugs automatically
spans several decades, and has produced 
a wide array of diverse tools such as
static analyzers, symbolic execution engines, and fuzzers---to mention just a few.
In contrast to this long and successful history of developing bug-finding tools,
there still is a somewhat limited agreement about how to
rigorously evaluate and compare their bug-finding capabilities
in realistic settings. %

In the last few years, to address this conspicuous gap,
we have seen several proposals of \emph{ground-truth benchmarks}:
curated collection of real programs including known bugs~\cite{magma}
or seeded with synthetic bugs~\cite{lava,bugsynth},
complete with detailed information about the bugs' location,
triggering inputs, and other fundamental characteristics.
Ground-truth benchmarks have been instrumental
in improving the rigor and thoroughness
of bug-finding tools---es\-pe\-cial\-ly those that generate test inputs
using symbolic execution or fuzzing, which are the benchmarks' usual primary focus.
While the usefulness of ground-truth benchmarks is undeniable,
extending a benchmark with additional bugs and programs---%
not to mention creating a new domain-specific benchmark from scratch---%
remains a complex and time-consuming endeavor.

In this paper, we explore a \emph{complementary} approach to
building ground-truth benchmarks,
where we automatically generate from scratch programs with seeded bugs.
The idea of constructing programs to be used as test inputs (PUTs: programs under test)
has been successfully used for other purposes,
such as to detect semantic compiler bugs that result in incorrect compilation~\cite{csmith}.

Our technique, which we call \hp,
builds programs starting from a seed that consists of a simple block
that fails when executed;
this represents a seeded bug.
Then, it repeatedly grows the program
by adding features (branching, looping, and so on)
that make it larger and more complex to test. \hp is highly configurable:
the user can choose aspects such as
how many programs to generate,
which syntactic features they should include, 
and the range of variability of their branching conditions.
Clearly, there is no a priori guarantee
that the synthetic PUTs generated by \hp
are representative of real-world bugs.
However, a fully synthetic approach also has clear advantages over manually curated collections:
since the whole process is automatic and customizable,
producing new benchmarks collecting programs with specific characteristics
is inexpensive.
In addition, \hp's PUTs come
with precise information about the bug location and any bug-triggering inputs.
Thus, they can supplement the programs in curated ground-truth benchmarks
to better evaluate the capabilities of bug-finding tools according to metrics
such as number of discovered bugs and bug detection time,
as well as to investigate which syntactic features of the faulty programs are more amenable
to which bug-finding tools.

After discussing \hp's design and implementation in \autoref{sec:methodology},
in \autoref{sec:exp-design}
we design some experiments where we generated hundreds of PUTs
with bugs using \hp, and we ran
three popular, mature bug-finding tools%
---\AFL, \CBMC, and \klee---%
on these PUTs.
Our goal is demonstrating that
\hp can generate bugs with diverse characteristics,
which can challenge different capabilities of bug-finding tools
and can usefully complement the programs in ground-truth benchmarks.
To this end, we follow Roy et al.~\cite{bugsynth}'s
description of the features of ``ecologically valid'' bugs,
and analyze whether \hp can generate bugs that are fair,
reproducible, deep, and rare,
and that can exercise the different capabilities of common bug-finding techniques.
The high-level summary of the experiments,
which we detail in \autoref{sec:exp-results},
confirms that \hp
is capable of generating ``interesting''
buggy programs that share some characteristics
with those of benchmarks.
Thus, \hp can support flexible empirical analysis
of the capabilities of the various bug-finding tools
in a way that complements and extends
what is possible using manually-curated benchmarks.

\paragraph{Contributions}
This paper makes the following contributions:

\begin{itemize}
\item \hp, a configurable technique to automatically generate PUTs
  with certain characteristics and seeded bugs.

\item An implementation of the \hp technique in a tool---also named \hp.

\item An experimental evaluation of \hp that demonstrates its ability to
  generate bugs with characteristics comparable to ``ecologically valid'' ones~\cite{bugsynth},
  which exercise from different angles the capabilities of
  bug-finding tools.
\end{itemize}
The prototype implementation of \hp is available in a public repository \cite{repository}.

\paragraph{Organization}
The rest of the paper is organized as follows.
\autoref{sec:related} discusses the main related work
in the development of benchmarks of bugs,
as well as bug-finding techniques and tools.
\autoref{sec:methodology}
describes the \hp technique and its current implementation
as a tool with the same name that generates programs in C.
\autoref{sec:exp-design}
introduces the paper's research questions,
and the experiments that we carried out to answer the questions.
\autoref{sec:exp-results}
presents the results of the experiments,
and how they address the research questions.
Finally, \autoref{sec:discussion}
concludes with a summary and discussion
of the paper's contributions.

\section{Related Work}
\label{sec:related}

We discuss related work in two areas:
benchmarks of bugs to evaluate bug-finding tools (\autoref{sec:rw:benchmarks}),
and the main techniques and tools to find bugs and vulnerabilities in programs (\autoref{sec:rw:tools}).
Consistently with the paper's main focus,
we principally consider techniques and tools that work on programs written in the C programming language used for systems programming.

\subsection{Benchmarks of Bugs}
\label{sec:rw:benchmarks}

Different applications of program analysis, including different approaches to test-case generation,
use different benchmarks, consistent with the goals of the program analysis evaluated using the benchmark.
Here, we focus on \emph{extensible} benchmarks to evaluate the \emph{bug-finding} capabilities
of test-case generation frameworks (for brevity, testing framework).

\begin{table}[!tb]
  \centering
  \begin{tabular} {@{\extracolsep{\fill}}l|ll}
    \toprule
    \multicolumn{1}{c}{} & \multicolumn{1}{c}{\textbf{Organic PUTs}} & \multicolumn{1}{c}{\textbf{Synthetic PUTs}} \\
    \cmidrule(lr){2-3}
    \multirow{3}{*}{\textbf{Organic bugs}} & FuzzBench &  \\
				& \underline{MAGMA} &  \\	
				& CGC, Test-Comp, SV-Comp (datasets) &  \\[5mm]	
    \multirow{3}{*}{\textbf{Synthetic bugs}} & \underline{LAVA} & CSmith \\
    	            & \underline{Apocalypse} &  \textbf{\underline{\hp}}\\	
            
    \bottomrule
  \end{tabular}
		\caption{Classification of evaluation benchmarks according to whether they consist of organic or synthetic bugs within organic or synthetic programs (PUTs). Underlined systems support the automatic generation of new benchmarks by seeding bugs into existing programs.}
		\label{tab:eval_class}
\end{table}

\autoref{tab:eval_class} shows a natural classification in terms of the origin of programs and their bugs,
and displays the category several well-known benchmarks belong to.
A \emph{program} included in a benchmark can be \emph{organic} or \emph{synthetic}.
The \emph{bugs} of a benchmark's PUTs can also be \emph{organic} or \emph{synthetic}.

\paragraph{Organic programs}
An \emph{organic program} is one that was designed and implemented by human programmers,
and hence reflects the characteristics of real-world programs (or at least a sample of them).
For this reason, many existing benchmarks are based on organic PUTs.
For example, the International Competition on Software Testing (Test-Comp)~\cite{test-comp} is a comparative evaluation of automatic tools for software test generation, which
uses benchmarks consisting of C programs equipped with testing objectives
(such as coverage, and bug finding).
Similar benchmarks are used by the Competition on Software Verification (SV-Comp)~\cite{sv-comp}.
Another example is the CGC dataset,
which collects about 300 small manually-written programs produced for the Darpa Cyber Grand Challenge~\cite{cgc};
for each bug in the programs, the CGC also includes a triggering input.

Google's FuzzBench is an open benchmarking platform and service~\cite{fuzzbench} based on open source programs. FuzzBench has been useful both in the industrial and the academic fields---both to evaluate the capabilities of fuzzing frameworks and to identify their limitations and own bugs. %

Organic benchmarks exist also for other programming languages,
such as the DaCapo benchmarks~\cite{dacapo_bench}
and Defects4J~\cite{defects4j}
for the Java programming language.

\paragraph{Synthetic programs}
In contrast, a \emph{synthetic program}
is one that is generated automatically from a set of templates, rules, or heuristics.

CSmith~\cite{csmith} is a program generator mainly employed for validating compilers through differential testing~\cite{differential-testing}. It has been used to find several security problems in popular compiler frameworks~\cite{compiler-fuzzing,csmithedge}, including GCC~\cite{gcc} and LLVM~\cite{llvm}.
Timotej and Cadar~\cite{putgen-sym-exec} applied a similar combination of grammar-based program generation and
differential testing
in order to find bugs in symbolic execution engines.
While tools such as CSmith could be used to build benchmarks that challenge testing frameworks,
they are most directly useful for \emph{differential} testing,
where the goal is comparing the behavior of different versions of a compiler.
\hp revisits some of the ideas behind tools like CSmith (in particular, grammar-based program generation)
so that they are directly applicable to generate PUTs with seeded bugs. 
Differently from CSmith, 
\hp can also produce a triggering input for each buggy program it generates,
which serves as the ground truth to assess and compare the capabilities of different 
bug-finding tools.

\paragraph{Organic bugs}
An \emph{organic bug} is one that occurred ``in the wild'', and hence comes from
a program's actual development history.
Just like organic programs, organic bugs have the clear advantage of being realistic.
In fact, the majority of current systems for the evaluation of testing frameworks consist of organic PUTs and organic bugs.
The MAGMA benchmark~\cite{magma} can extend the usability of such ``fully organic'' benchmarks by
performing ``forward-porting'' of real bugs to recent version of the target PUT.
This way, a historically relevant bug can still be reproduced (and tested for) in up-to-date setups.
Still, applying MAGMA to new bugs and new PUTs requires substantial manual effort.

\paragraph{Synthetic bugs}
Seeding \emph{synthetic} bugs into an existing program
has become an increasingly popular approach to generate large benchmarks of bugs, thanks to its scalability
compared to manual selection and curation.
The Large-scale Automated Vulnerability Addition (LAVA) dataset~\cite{lava}%
---commonly used to compare fuzzing frameworks---%
consists of synthetic bugs seeded into existing programs.
LAVA's bug injection is based on the PANDA dynamic analysis platform \cite{panda}, built on top of the QEMU emulator \cite{qemu}.
First, an analysis of the target program identifies dead, unused, and available (DUA) bytes of the input,
which can be altered (``fuzzed'') without affecting the program's behavior.
Then, LAVA seeds vulnerabilities, such as buffer overflows or other kinds of inconsistent memory access, that are triggered when an execution accesses the DUA bytes.
% In order to work on real-world programs, LAVA's implementation has some limitations
% such as the location of the seeded bugs,
% which can only be chosen within the code that has been reached in a previous execution.

Apocalypse \cite{bugsynth} is a bug injection system similar to LAVA and based on synthetic bugs and symbolic execution. It generates and seeds into existing programs bugs with specific requirements (some of which we describe in \autoref{sec:exp-design} in relation to our experiments). Apocalypse was experimentally evaluated to show it can
generate seeded bugs with characteristics comparable to organic ones. In \autoref{sec:exp-results}, we will assess the PUTs generated by \hp using 
several of the same metrics.

In order to work on real-world programs, LAVA and Apocalypse incur some limitations.
First, one cannot seed bugs at arbitrary locations but
only at those that have been reached in a previous execution.
Second, since they rely on symbolic execution to discover triggering inputs
for the seeded bugs, it may be practically hard to find such triggering inputs
for bugs that are nested very deeply into the program's control flow structure.
Since \hp builds PUTs with seeded bugs from scratch, it does not incur these limitations
and can generate programs with arbitrarily complex nesting structure 
(as we demonstrate in \autoref{sec:results:deep-rare}).

Ferrer et al.'s work~\cite{benchgen-java}
is an example of fully synthetic benchmarks
(consisting of synthetic bugs and synthetic PUTs) for the Java programming language.
Their main goal is generating programs where every branch is reachable to serve as ground truth
when evaluating the branch-coverage capabilities of testing frameworks.

Mutation testing is another approach based on injecting synthetic bugs in organic programs~\cite{mutation-testing,mutation_testing_put_gen}.
The original goal of mutation testing was to measure the bug-detection capabilities
of a test suite: the more ``mutants'' (i.e., variants of program with injected bugs)
trigger failures in the test suite, the more comprehensive the test suite is~\cite{pezze_book}.
More recently, mutation testing ideas have been applied to different dynamic analysis techniques, such
as fault localization~\cite{metallaxis,mutation-fault}.
As a bug-injection technique~\cite{mutation-generation},
mutation testing suffers from the problem of \emph{equivalent mutants},
which occur when a mutation does not alter a program's behavior,
and hence the mutant does not actually have a bug;
a number of approaches have tried to address this problem~\cite{equivalent-detection,equivalent-2,equivalent-3}.

It is also interesting to consider which metrics are supported
by the benchmarks.
The most common ones are number of detected bugs, detection time,
and maximum code coverage achieved during testing;
these are easily applicable to all benchmarks.
In addition, one may want to relate the syntactic features of the buggy programs
to the capabilities of the bug-finding tools;
\hp's approach supports this kind of experiments,
since it can generate batches programs with similar characteristics
(e.g., nesting structure or kinds of statements).

% An alternative classification for testing framework benchmarks is based on the employed evaluation criteria. 
% The most common ones are the detection time for a particular bug, the number of detected bugs in the benchmark and the code coverage testing frameworks achieve during program execution, measured in terms of number of lines or branches visited in the PUT control flow graph.
% Every benchmark described in this section supports the first two mentioned criteria, while coverage measuments can be easily incorporated at compilation time. HyperPUT, in addition, can also evaluate testing frameworks depending on the structure of the produced PUT, as described in Section \ref{sec:methodology}.

\subsection{Bug-finding Tools}
\label{sec:rw:tools}

A detailed discussion of the main techniques
used to find bugs in programs
is beyond the scope of the present paper;
we refer the interested readers to surveys~\cite{testingsurvey,symbexec_survey,symbexec_survey2}
and textbooks~\cite{AmmanOffutt,pezze_book}.
In this section, we briefly describe the bug-finding techniques and tools
that feature in our evaluation of \hp---which are also widely used outside of
research.

\paragraph{Fuzzing}
Fuzz testing (or \emph{fuzzing})
encompasses a broad spectrum of dynamic techniques
to generate program inputs \cite{fuzzing_survey,evaluating_fuzzing}.
It is widely used to find bugs in software;
Google, for instance, found thousands of security-related bugs in their software using fuzzing~\cite{gresearch}.
The key idea of fuzzing is to \emph{randomly mutate}
a known valid program input (the ``seed'')
to generate new inputs that may cause the program to crash or expose other kinds of vulnerabilities.
Fuzzers differ according to the kind of strategies they use to randomly mutate program inputs.
In particular, black-box fuzzers do not have access to the target program's control flow,
and hence can only generate new inputs independently of the program's structure.
In contrast, white-box fuzzers can take the program's control flow into account
in order to generate new inputs that exercise specific portions of the program.

American Fuzzy Lop (AFL) is
one of the most popular fuzzing frameworks for C programs.
It is a gray-box coverage-based fuzzer,
which means that some of its fuzzing strategies
are driven by coverage information about the analyzed program.
Originally developed by Zalewsky~\cite{afl},
different extensions of AFL%
---such as REDQUEEN \cite{rq}, AFLFast \cite{aflfast} and AFL++ \cite{aflpp}---%
have been introduced more recently and remain widely used.

\paragraph{Symbolic execution}

As the name suggests,
symbolic execution executes a program with \emph{symbolic} inputs,
which are placeholders for every possible valid inputs~\cite{symbexec_survey,symbexec_survey2}.
As it enumerates different execution paths,
symbolic execution builds \emph{path constraints}, which
are logic formula that encode each path's feasibility.
Then, a constraint solver such as Z3~\cite{z3}
determines which abstract paths are feasible, and generates matching concrete inputs.

Most modern implementations of symbolic execution perform
\emph{dynamic symbolic execution} (also called ``concolic'' execution),
which combines symbolic and concrete state in order to overcome some limitations
of symbolic execution (such as its scalability and applicability to realistic programs)~\cite{symbexec_limits2}.
EXE~\cite{exe} and DART~\cite{dart}
pioneered the idea of dynamic symbolic execution.
More recently, other tools perfecting and extending this technique
include KLEE~\cite{klee}, SAGE~\cite{sage}, S2e~\cite{s2e}, and Angr~\cite{angr}.
KLEE is one of the most widely used dynamic symbolic execution engines for C programs.
It is implemented on top of LLVM~\cite{llvm},
and %
has been successfully employed to find several bugs in production software, such as the
MINIX~\cite{minix} and BUSYBOX~\cite{busybox} tools.

Driller is a vulnerability discovery tool that combines symbolic execution and fuzzing~\cite{driller}.
When the latter fails to make progress, it uses the former to continue the exploration of new execution paths.
This approach is effective to improve code coverage,
and to test features such as cryptographic hash functions and random number generators,
which are notoriously difficult for approaches that are exclusively based on constraint solving.
The T-Fuzz fuzzer~\cite{tfuzz} applies program transformations in order to remove the conditions guarding some code blocks that are hard to reach. If a crash occurs in these code blocks, it then checks a posteriori whether the locations
are actually reachable in the original program. %

\paragraph{Model checking} 
\label{sec:modelchecking}

In a nutshell, model checking is verification technique for
finite-state models,
which can exhaustively check properties expressed in temporal logic
(including reachability properties, which can be expressed as assertions in the code) 
or find counterexamples when the properties do not hold in general~\cite{model-checking-ref}.

Since real-world programs are not finite state,
one needs to introduce some kind of finite-state abstraction in order to
be able to apply model checking to them.
A natural way of doing so is by \emph{bounding}
the program state to be within a finite (but possibly very large) range.
Then, model checking such a bounded abstraction is not equivalent
to verifying the original program, but can still be a very effective way of thoroughly \emph{testing}
the program and finding bugs.
In this paper, we experiment with the popular
CBMC~\cite{cbmc} bounded model checker for C programs.

An alternative classification for testing framework benchmarks is based on the employed evaluation criteria. 
The most common ones are the detection time for a particular bug, the number of detected bugs in the benchmark and the code coverage testing frameworks achieve during program execution, measured in terms of number of lines or branches visited in the PUT control flow graph.
Every benchmark described in this section support the first two mentioned criteria, while coverage measurements can be easily incorporated at compilation time. HyperPUT, in addition, can also evaluate testing frameworks depending on the structure of the produced PUT, as described in Section \ref{sec:methodology}.

\section{Methodology and Implementation}
\label{sec:methodology}

\hp builds arbitrarily complex PUTs by recursively applying parametric transformations
of different kinds 
to an initial simple program.

\subsection{Transformations}
\label{sec:transformations}

\begin{table}[!tb]
  \begin{tabular}{ll}
    \toprule
    \multicolumn{1}{c}{\textsc{transformation}} & \multicolumn{1}{c}{\textsc{code}} \\
    \midrule
    $\tr{IC}(v_1\colon \code{long}, v_2\colon \code{long}, T, E)$ &
\begin{lstlisting}[numbers=none, basicstyle=\ttfamily\tiny] 
if ($v_1$ == $v_2$) { $T$ } else { $E$ }
\end{lstlisting}
    \\ \cmidrule(lr){2-2}
    $\tr{SC}(s_1\colon \code{char*}, s_2\colon \code{char*}, T, E)$ & 
\begin{lstlisting}[numbers=none, basicstyle=\ttfamily\tiny]
if (strcmp($s_1$, $s_2$) == 0) { $T$ } else { $E$ }
\end{lstlisting}
  \\ \cmidrule(lr){2-2}
  $\tr{FL}(e\colon \code{long long}, B)$ & 
\begin{lstlisting}[numbers=none, basicstyle=\ttfamily\tiny]
for (long long j = 0; j < $e$; j++) { do_something(); } $B$
\end{lstlisting}                                                                                
    \\ \cmidrule(lr){2-2}
    $\tr{PC}(s\colon \code{char*}, n\colon \code{int}, B)$ &
\begin{lstlisting}[numbers=none, basicstyle=\ttfamily\tiny]
if (strlen($s$) < $n$) exit(0);
size_t l = 0, h = strlen($s$) - 1;
while (h >= l) { if ($s$[h] != $s$[l]) exit(0); h--; l++; } $B$
\end{lstlisting}                                         
    \\ \cmidrule(lr){2-2}
    $\tr{CC}(s\colon \code{char*}, c\colon \code{char}, n\colon \code{int}, T, E)$ &
\begin{lstlisting}[numbers=none, basicstyle=\ttfamily\tiny]
int count = 0;
for (int k = 0; k < strlen($s$); k++) { if ($s$[k] == $c$) count++; }
if (count == $n$) { $T$ } else { $E$ }
\end{lstlisting}
    \\ 
    \bottomrule
\end{tabular}
  \caption{\hp's transformations and the corresponding generated code. In a transformation, lowercase letters denote parameters and uppercase letters denote holes.}
  \label{tab:transformations}
\end{table}

A \emph{transformation} consists of a program \emph{template}
with (typed) parameters and holes.
When we \emph{apply} a transformation,
we choose concrete values for its parameters and holes.
A parameter can be replaced with any constant or variable of suitable type.
A hole is replaced by another snippet of code,
which can be given explicitly or as the result of nesting another transformation.
\autoref{tab:transformations} lists the transformations \hp currently supports,
together with the code they correspond to.
There are five main kinds of transformations:
\begin{description}
\item[\tr{IC} (integer comparison)] introduces a conditional that checks
  whether the two integer parameters $v_1, v_2$ are equal.
\item[\tr{SC} (string comparison)] introduces a conditional that checks
  whether the two string parameters $s_1, s_2$  are equal.
\item[\tr{FL} (for loop)] introduces a loop that iterates $e$ times
  (where $e$ is the transformation's integer parameter), and then
  executes code $B$. 
\item[\tr{PC} (palindrome check)] introduces a loop that checks whether
  the string parameter $s$ is a palindrome of length at least $n$; if it is, it executes code $B$.
\item[\tr{CC} (character counting)] introduces a loop that
  counts the number of occurrences of character parameter $c$ in string parameter $s$;
  if the count equals the integer parameter $n$, it executes code $T$; if not, it executes code $E$.
\end{description}

Let's present a few more details about transformation
\tr{IC}, as an example to illustrate how transformations work.
Transformation \tr{IC} consists of two parameters $v_1$ and $v_2$ and two holes $T$ and $E$.
The parameters denote two integer values or variables.
Then, the transformation introduces a conditional \code{if} that checks whether $v_1$ and $v_2$
have the same value.
If they have, 
$T$ executes; otherwise, $E$ executes.

\subsection{Transformation Sequences}
\label{sec:sequences}

More complex PUTs combine several transformations by nesting one inside another.
When we specify a \emph{sequence} of transformations, we can give a concrete value to
any transformation \emph{parameter} or use a fresh identifier.
In the latter case, \hp will instantiate the parameter
with a suitable random value (usually within a range)---for every PUT generated from the transformation sequence.
For example, the expression $\tr{IC}$ $(\code{atoll(argv[1])}$,$ \beta$, \code{assert 0 == 1}, $\code{exit(0)})$,
where $\beta$ is a fresh  identifier,
denotes a conditional  that checks whether the first command-line argument
\code{argv[1]}, when interpreted as an integer,
is equal to a random integer value; if it is, the program fails (\code{assert 0 == 1}),
otherwise, it exits normally (\code{exit(0)}).

We can also use fresh identifiers, instead of concrete code snippets, for \emph{holes},
to denote that the next transformation in the sequence will instantiate the hole.
In other words, this is just a notational shorthand that helps readability by avoiding nesting transformations
explicitly.
For example, the sequence of two transformations
\begin{equation}
  \tr{SC}(\code{argv[2]}, \code{"hello"}, \code{;}, E)
  \ 
  \tr{IC}(\code{atoll(argv[1])}, 69, \code{assert 0 == 1}, \code{return(0)})
  \label{eq:sc-ic-example}
\end{equation}
nests an integer comparison inside the else branch of a string comparison,
and thus it is equivalent to the explicitly nested expression
\begin{equation*}
  \tr{SC}\Bigl(\code{argv[2]}, \code{"hello"}, \code{;}, \bigl( \tr{IC}(\code{atoll(argv[1])}, 69, \code{assert 0 == 1}, \code{return(0)}) \bigr)\Bigr)
\end{equation*}
and determines the PUT in \autoref{fig:sc-ic-example}.

\begin{figure}[!htb]
  \centering
\begin{lstlisting}
int main(int argc, char** argv) {
  if (strcmp(argv[2], "hello") == 0)
    ;(*\label{l:sc-ic:leaf1}*)
  else {
    if (atoll(argv[1]) == 69)
      assert 0 == 1;(*\label{l:sc-ic:leaf2}*)
    else
      return 0;(*\label{l:sc-ic:leaf3}*)
  }
}
\end{lstlisting}
  
  \caption{Specification of a PUT that combines transformations \tr{SC} and \tr{IC} as in \eqref{eq:sc-ic-example}.}
  \label{fig:sc-ic-example}
\end{figure}

\autoref{fig:sc-ic-example}
also shows that \hp inserts
the code generated by applying a sequence of transformations
into a template main function,
so that the PUT is a complete program.
\hp also automatically generates
boilerplate code%
---such as library includes, and checks that the required command-line arguments are indeed present---%
that makes PUTs syntactically correct programs.
For simplicity, \autoref{fig:sc-ic-example} and
all other PUTs shown in the paper omit
this boilerplate code.

\paragraph{Reaching inputs}

The structure of every transformation
also suggests which values of
the transformation's parameters
determine an execution of the resulting PUT
that reaches code in any of the transformation's holes.
For example, hole $T$ in transformation \tr{IC} executes
for any $v_1 = v_2$;
hole $B$ in transformation \tr{FL} always executes;
hole $T$ in transformation \tr{CC} executes if $s$ includes $n$ occurrences of characters $c$;
and so on.
Based on the transformations' structure and how they are combined,
\hp outputs, for every PUT it generates,
values for all variables used in any transformation's parameters
that reach any of the PUT's holes.
In \autoref{fig:sc-ic-example}'s example,
there are two variables \code{argv[1]} and \code{argv[2]},
and three leaf holes at lines~\ref{l:sc-ic:leaf1}, \ref{l:sc-ic:leaf2}, and \ref{l:sc-ic:leaf3};
\hp determines that the inputs $\langle\code{""}, \code{"hello"}\rangle$, $\langle\code{"69"}, \code{""}\rangle$, and $\langle\code{""}, \code{""}\rangle$
respectively reach each of the leaves.

\subsection{Implementation Details}
\label{sec:implementation}

We implemented the \hp technique in a tool with the same name.
The tool is implemented
in a combination of C
(for the core program-generation functionalities),
Python (front end and connection of the various modules),
and Bash scripts (to run batches of experiments).

The user input to \hp consists of
a sequence of transformations specified as described in \autoref{sec:sequences}, %
and a number of PUTs to be generated.
\hp's \emph{front end} processes this input
and passes the information to the \emph{generator engine},
which takes care of generating PUTs
by applying the transformation sequences,
embedding the resulting code into a main function to build a complete program,
and also recording a reaching input for every generated PUT.

\paragraph{Extensibility}
\hp is extensible with new transformations.
However, as we demonstrate in \autoref{sec:exp-results},
the current selection of transformations is already sufficient
to generate a large number of ``interesting'' PUTs,
which can challenge different test-case generators
and share some characteristics with the programs in widely used test-case generation benchmarks.

In principle, \hp's pipeline could also generate PUTs in programming languages
other than C.
To this end, one should extend it with transformations
that generate valid snippets of code in other programming languages.

\section{Experimental Design}
\label{sec:exp-design}

The experimental evaluation of \hp addresses the following research questions:
\begin{description}
\item[RQ1:] Can \hp generate bugs that are \emph{fair}?
\item[RQ2:] Are the bugs generated by \hp \emph{reproducible}?
\item[RQ3:] Can \hp generate bugs that are \emph{deep} and \emph{rare}?
\item[RQ4:] Can \hp generate diverse programs that exercise
  different \emph{capabilities} of bug-finding techniques?
\end{description}

This section describes the experiments we designed to answer these research questions.
Our experimental design is after Roy et al.~\cite{bugsynth}'s,
modified to suit our goal of evaluating the characteristics of \hp's
synthetic PUTs.

\subsection{Testing Frameworks}
To assess the characteristics of the bugs generated by \hp,
we ran several testing frameworks on the generated PUTs
and determined which bugs each framework could uncover.

We used testing frameworks implementing different
bug-finding techniques for C programs:
\begin{itemize}
\item \AFL~\cite{afl} is a popular grey-box fuzzer, which combines random generation
  of input and coverage metrics.

\item \CBMC~\cite{cbmc} is a bounded model checker for C/C++ programs.
  Bounded model-checking exhaustively explores a program's state-space up
  to a finite size bound, checking for the violation of
  basic correctness properties (such as memory safety)
  and assertions within this explored space.

\item \klee~\cite{klee} is a state of the art dynamic-symbolic execution engine.
  Dynamic-symbolic execution is a white-box testing technique, which uses constraint
  solving to generate inputs that lead to exploring new paths in the PUT.

\end{itemize}

These tools offer numerous configuration options;
\autoref{tab:tf-configs}
lists the configurations that we used in the experiments.
We deploy each tool in two configurations:
we first execute it with its first configuration;
if it fails to find a bug before the timeout expires,
we execute it again on the same PUT with its second configuration 
(using any remaining time).
For brevity, henceforth we use the expression ``we run $X$ on a program $P$''
to mean ``we run the testing framework $X$ using sequentially the two configurations in~\autoref{tab:tf-configs} on $P$''.

\begin{table}[!h]
  \centering
  \begin{tabularx}{1.0\linewidth}{l l X}
    \toprule
    \multicolumn{1}{c}{\textsc{id}} & \multicolumn{1}{c}{\textsc{framework}} & \multicolumn{1}{c}{\textsc{configurations}} \\
    \midrule
    \multirow{2}{*}{$A$} & \multirow{2}{*}{\AFL}
                    & \verb+afl-clang-fast+ with options CMPLOG~\cite{complog}, LAF~\cite{laf}, MOpt~\cite{mopt}
    \\
                    & & \verb+afl-clang-fast+ with default options
    \\
    \cmidrule(lr){1-2}
    \multirow{2}{*}{$C$} & \multirow{2}{*}{\CBMC} & automated bounded loop unwinding
    \\
                                                  & & loop unwinding with bound 10
    \\
    \cmidrule(lr){1-2}
    \multirow{2}{*}{$K$} & \multirow{2}{*}{\klee} & symbolic arguments, random state search, LLVM optimization
    \\
                                    && symbolic arguments, default options
    \\
    \bottomrule
  \end{tabularx}
  \caption{Configurations of the testing tools used in the experiments.
    Each row specifies two configurations for a testing tool
    in terms of the used options.}
  \label{tab:tf-configs}
\end{table}

\subsection{Experimental Subjects}
\label{sec:subjects}

We generate PUTs in batches,
where each batch runs \hp
with a sequence of $n \geq 1$ transformations:
\begin{equation}
  \tiny
  \tr{T}_1(p_{1,1}, p_{1,2}, \ldots, H_{1,1}, \ldots)
  \;
  \tr{T}_2(p_{2,1}, p_{2,2}, \ldots, H_{2,1}, \ldots)
  \;
  \ldots
  \;
  \tr{T}_n(p_{n,1}, p_{n,2}, \ldots, \code{fail()}, \ldots)
  \label{eq:sequence-subjects}
\end{equation}
and a matching sequence of actual parameters $p_{1,1}, p_{1,2}, \ldots, p_{n,1}, p_{n,2}, \ldots$.
Each transformation $T_k$ in \eqref{eq:sequence-subjects}
is one of \tr{IC}, \tr{SC}, \tr{FL}, \tr{PC}, and \tr{CC}
listed in \autoref{tab:transformations}.
In the experiments, we always nest into the ``then''
hole $T$ of conditional transformations \tr{IC} and \tr{SC};
therefore, all ``else'' holes $E$ are simply filled with a ``skip'' snippet
that does nothing.
Snippet \code{fail()} indicates code that triggers a crashing bug when executed;
for example, an assertion failure \code{assert 0 == 1} or an out-of-bound error \code{int a[3]; a[4] = 0}.
In our experiments, we always add the snippet \code{fail()}
in the innermost transformation $\tr{T}_n$. 

Each actual parameter $p_j$ is either a random constant of the appropriate type
(chosen within a limited range)
or
\begin{enumerate*}[label=\emph{\roman*})]
\item \code{argv[i]} (for $\code{i} \geq 1$) for parameters of type \code{char*};
\item \code{atoll(argv[i])} (for $\code{i} \geq 1$) for parameters of integer type (\code{int}, \code{long}, \code{long long}).
\end{enumerate*}
More precisely, parameters
$v_1$ in transformation \tr{IC},
$s_1$ in transformation \tr{SC},
and $s$ in transformations \tr{PC} and \tr{CC}
are always instantiated with a command-line argument;
all other parameters are chosen as random constants within a small range.
\autoref{tab:parameter-ranges} shows the actual ranges for the randomly chosen parameters
in each transformation in the batches that we used in the experiments.
For example, every instance of \tr{IC} uses an integer between 0 and 255 as its second parameter $v_2$.
We introduce the described restrictions on the choice of parameters
so as to generate PUTs of homogeneous characteristics,
where the number and kinds of transformations used to generate them
are the primary determinant of their complexity.
These constraints also ensure that, in every generated PUT, 
\begin{enumerate*}[label=\emph{\roman*})]
\item there is exactly one bug;
\item there is (at least one) program input that triggers the bug.
\end{enumerate*}
\hp's reaching input for the unique bug's location
is thus also the triggering input that ensures that the bug is executable.

\begin{table}[!bt]
  \centering
  \begin{tabular}{lrrl}
  \toprule
    \textsc{batch} & $n$ & \textsc{\#puts} & \textsc{inputs used as parameters} $v_1, s_1, s$ \\
    \midrule
    $B_1$ & 1 & 10  & $\code{argv[1]}$ \\
    $B_2$ & 2 & 45 & $\code{argv[1]}, \code{argv[2]}$ \\
    $B_{10}$ & 2--10 & 200 & $\code{argv[1]}, \ldots, \code{argv[10]}$ \\
    $B_{100}$ & 100 & 100 & $\code{argv[1]}, \ldots, \code{argv[100]}$ \\
    $B_{1000}$ & 1000 & 100 & $\code{argv[1]}, \ldots, \code{argv[1000]}$ \\
    \bottomrule
\end{tabular}
\caption{List of the batches of PUTs used in \hp's experimental evaluation to answer RQ1, RQ2, and RQ3.
  For each \textsc{batch}, the table lists the number $n$ of \emph{transformations} used to generate each PUT in the batch,
  the number \textsc{\#puts} of different PUTs in the batch, and the command-line input arguments used
  as parameters in the transformations.}
\label{tab:batches}
\end{table}

\begin{table}[!tb ]
  \centering
  \begin{tabular}{llrr}
    \toprule
    \multicolumn{1}{c}{\textsc{transformation}} & \multicolumn{1}{c}{\textsc{parameter}}  & \multicolumn{1}{c}{\textsc{min}} & \multicolumn{1}{c}{\textsc{max}}
    \\
    \midrule
    \tr{IC} & $v_2$ & \code{0} & \code{255}
    \\
    \tr{SC} & $s_2$ & \code{"0"} & \code{"255"}
    \\
    \tr{FL} & $e$ & \code{0} & \code{255}
    \\
    \tr{PC} & $n$ & \code{1} & \code{20}
    \\
    \tr{CC} & $n$ & \code{1} & \code{20}
    \\
    \bottomrule
  \end{tabular}
  \caption{Range of values, between a \textsc{min}imum and a \textsc{max}imum value,
    for the \textsc{parameter}s of the \textsc{transformation}s in \autoref{tab:transformations}
    used in the experiments.}
  \label{tab:parameter-ranges}
\end{table}

\begin{figure}[!hbt]
\centering
\includegraphics[width=.8\linewidth]{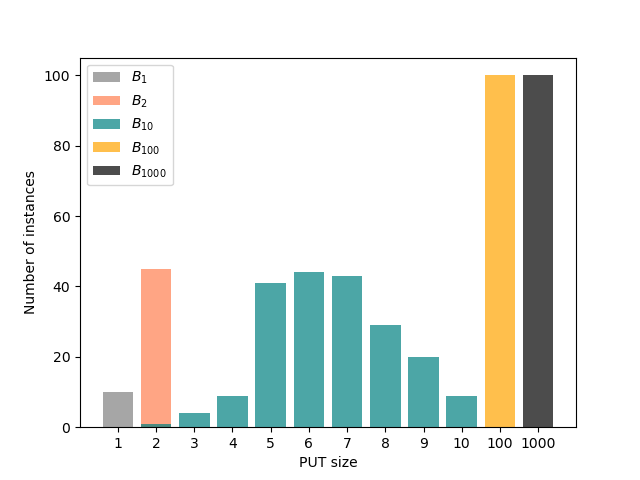}
\caption{Distribution of size (in number of transformations)
  of the PUTs used in the experimental evaluation.}
\label{fig:puts_distribution}
\end{figure}

\paragraph{Batches}
For the experiments with \hp to answer RQ1, RQ2, and RQ3,
we generated a total of 455 PUTs
in 5 batches.
\autoref{tab:batches} outlines the characteristics of each batch.
\begin{description}
\item[Batch $B_1$] includes 10 PUTs, each consisting of a single transformation.

\item[Batch $B_2$] includes 45 PUTs, each consisting of two different transformations.

\item[Batch $B_{10}$] includes 200 PUTs, each consisting of between 2 and 10 transformations
  (possibly with repetitions), with the transformations and the actual length chosen randomly.
  More precisely, this batch includes:
  \begin{enumerate*}[label=\emph{\roman*})]
  \item 1 PUT consisting of 2 transformations;
  \item 4 PUTs consisting of 3 transformations;
  \item 9 PUTs consisting of 4 transformations;
  \item 41 PUTs consisting of 5 transformations;
  \item 44 PUTs consisting of 6 transformations;
  \item 43 PUTs consisting of 7 transformations;
  \item 29 PUTs consisting of 8 transformations;
  \item 20 PUTs consisting of 9 transformations;
  \item 9 PUTs consisting of 10 transformations.
  \end{enumerate*}

\item[Batch $B_{100}$] includes 100 PUTs, each consisting of exactly 100 transformations
  (possibly with repetitions) chosen randomly.
  
\item[Batch $B_{1000}$] includes 100 PUTs, each consisting of exactly 1000 transformations
  (possibly with repetitions) chosen randomly.
\end{description}
Henceforth, $B$ denotes the union of all batches $B_1 \cup B_2 \cup B_{10} \cup B_{100} \cup B_{1000}$.
\autoref{fig:puts_distribution} overviews the distribution of all PUTs in $B$.

For the experiments with \hp to answer RQ4,
we generated another 60 PUTs
in 6 batches $B_{\tr{IC}}, B_{\tr{SC}}, B_{\tr{FL}}, B_{\tr{PC}}, B_{\tr{CC}}, B_{\star}$.
For each transformation $T$ among \tr{IC}, \tr{SC}, \tr{FL}, \tr{PC}, and \tr{CC},
batch $B_{T}$ consists of 10 PUTs $P_T^1, \ldots, P_T^{10}$.
Each PUT $P_T^m$ corresponds to the sequence of transformations 
\begin{equation}
  \footnotesize
  T(p_{1,1}, p_{1,2}, \ldots, H_{1,1}, \ldots)
  \;
  T(p_{2,1}, p_{2,2}, \ldots, H_{2,1}, \ldots)
  \;
  \ldots
  \;
  T(p_{m,1}, p_{m,2}, \ldots, \code{fail()}, \ldots)
  \label{eq:sequence-subjects-rq4}
\end{equation}
with $m$ transformations, all equal to $T$.
In other words, $B_{T}$ consists of increasingly long sequences of the same transformation $T$
repeated multiple times.
Similarly, batch $B_\star$ consists of 10 PUTs $P_\star^1, \ldots, P_\star^{10}$;
each PUT $P_\star^m$ corresponds to the sequence of transformations \eqref{eq:sequence-subjects},
with $n=m$ transformations,
each transformation randomly chosen  (possibly with repetitions)
among \tr{IC}, \tr{SC}, \tr{FL}, \tr{PC}, and \tr{CC}.

\paragraph{Bug categories}

In principle, \hp's seeded bugs can match any kind of errors; in our experimental evaluation, however, all seeded bugs are simply assertion failure. This is consistent with our research questions and experimental protocol, which follow how benchmarks of bugs are normally employed: to assess the capabilities of bug-finding tools on challenging, diverse buggy programs and \emph{reach} a bug's triggering location, independently of the nature and origin of the bug.

% While HyperPUT supports several categories of different bugs, such as buffer and heap overflows, pointer related problems, uninitialized memory accesses and semantic bugs, for our practical research evaluation, we focused on reachable assertion failures. The reason is that our research questions don't depend on the specific category of bug and they assume that the execution of the bug will cause a crash or an abnormal behaviour of the PUT that can be detected by the testing framework.

\subsection{Experimental Setup}
\label{sec:setup}

We ran all experiments on an Intel\textregistered~Core\textsuperscript{TM}~i5 machine with 2 cores and 8~GB of RAM running Ubuntu 18.04 Bionic,
LLVM~6.0.1, \AFL~2.68c, \CBMC~5.10, and \klee~2.1.

Every PUT generated by \hp
accepts command-line arguments as input for its \code{main} function.
This is the only input that a testing tool controls when testing a PUT.
For example, when running \klee, the command line argument array \code{argv}
is instrumented with \code{klee_make_symbolic}, and the rest of the PUT is unmodified.

Each experiment runs one of the tools 
in \autoref{tab:tf-configs} on a PUT %
with a timeout of 1~hour.
The outcome is success if the testing framework
successfully generates command-line inputs that trigger the \code{fail()}
injected bug in the PUT.
To accommodate fluctuations due to the operating system's nondeterministic scheduling,
as well as in possible randomization used by the testing frameworks,
we repeat each experiment four to ten times, and report the average wall-clock
running time as the experiment's duration.
The outcome is success if at least one of the repeated runs is successful (i.e., it triggers the bug).

\subsection{RQ1: Fairness}
\label{sec:design:fair}

A collection of bugs is \emph{fair}
if state-of-the-art bug detection techniques,
especially those that are widely used in practice, 
can discover the bugs with reasonable effort;
and if it is not strongly biased in favor or against any one detection technique.
For a PUT-generation system like \hp,
\emph{fairness} means that it should be capable of generating
bugs with a broad spectrum of ``detection hardness''%
---from simple to very challenging to discover. 

To demonstrate \emph{fairness}, we ran each of the tools 
\AFL, \CBMC, and \klee
on all PUTs in $B$.
We then analyzed which tools 
were successful in triggering the bugs in the PUTs
within the timeout.

\subsection{RQ2: Reproducibility}
\label{sec:design:reproducible}

A bug is \emph{reproducible} if
there is a known input that consistently triggers the bug.
For a PUT-generation system like \hp,
\emph{reproducibility} also entails that
the PUTs compile without errors and
do not rely on any undefined behavior of the C language.
All PUTs generated by \hp
come with an input that triggers their unique bug.
To assess \emph{reproducibility}, we ran
each PUT generated in the experiments with the triggering input,
and checked whether the bug was triggered as expected.

\hp generates PUTs
that should be syntactically and semantically correct.
To confirm this,
we compiled each PUT generated in the experiments
using both
GCC (with options \code{-O0 -Wall} and \code{-O1 -Wall})
and
LLVM (with options \code{-O0 -Wall} and \code{-O1 -Wall}),
and checked that:
\begin{enumerate*}[label=\textit{\roman*})]
\item both compilations succeeded without errors; and
\item both compiled versions behaved in the same way---namely, they fail when executed with the triggering input.
\end{enumerate*}
To detect the potential presence of undefined behavior,
we also checked every generated PUT using
LLVM's Undefined Behavior Sanitizer \cite{ubsan},
a compiler instrumentation that can detect several instances of undefined behavior.

\subsection{RQ3: Depth and Rarity}
\label{sec:design:deep-rare}

Depth and rarity are two different ways of assessing the ``hardness'' of a bug
for bug-detection techniques.

\subsubsection{Depth}
\label{sec:design:deep}

A bug is \emph{deep} if triggering it
requires to follow a long sequence of statements and branches.
For a PUT-generation system like \hp,
bug \emph{depth} depends on the structure and complexity of the PUTs themselves.
To determine whether \hp's bugs are \emph{deep},
we measured the following on every PUT in batch $B$ generated in the experiments:
\begin{itemize}
\item The cyclomatic complexity of the PUT.\footnote{Measured using CCCC \cite{cccc} and PMCCABE \cite{pmccabe} open source tools.}
  Cyclomatic complexity~\cite{cc} is a static measure of complexity of
  a program's branching structure,
  which counts the number of distinct simple execution paths a program has.

  In order to assess the complexity of \hp's PUTs compared to that of programs in other benchmarks,
  we compare the cyclomatic complexity of PUTs in $B$ to that of programs in CGC~\cite{cgc} and LAVA-1~\cite{lava}.
  Note that the PUTs generated by \hp consist of a single main function, but programs in other
  benchmarks usually consist of several different functions;
  thus, we measure the cyclomatic complexity of each function in the programs 
  in isolation, and report statistics about their distribution in each benchmark.
  We only measure the cyclomatic complexity of functions in the actual PUTs,
  not in any external library that is used by the PUTs.
  
\item The length (in number of instructions executed at runtime)
  of the execution path that goes from the PUT's entry to the bug-triggering statement,\footnote{Measured using the profiling tool Cachegrind~\cite{valgrind_thesis}.}
  when the PUT is executed with a triggering input.
  Path length is a dynamic measure of how deep a bug is within a path that triggers it.
  Similarly to cyclomatic complexity, 
  we compare the path length of bugs in $B$
  to that of bugs in benchmark LAVA-1.
\end{itemize}

\subsubsection{Rarity}
\label{sec:design:rare}

A bug is \emph{rare} if it is only triggered by a small fraction of all possible program inputs.

To determine whether \hp's bugs are \emph{rare},
we ran \klee on each buggy PUT with a timeout of 1 hour
and measure the following:
\begin{itemize}
\item The number $f$ of test cases generated by \klee before first triggering the bug.

\item The number $t$ of test cases, among those generated within the timeout, that trigger the bug.
\end{itemize}
These measures give an idea of how sparse the bug-triggering
inputs are in the space of all inputs that are generated by a systematic strategy.

In order to be able to compare \hp's measures of rarity with
those of other benchmarks',
we only considered PUTs in batch $B_{\geq 6}$
for this experiment.
Batch $B_{\geq 6}$ consists
of the 19 PUTs in $B_{10}$ with 6, 7, 8, 9, or 10 transformations
that \klee can discover within the 1-hour timeout.
We exclude PUTs with a much smaller or much larger input space,
where these metrics would be arguably less robust and less indicative of rarity.
We also exclude PUTs whose bugs \klee cannot uncover,
as the measures $f$ and $t$ are essentially undefined in these cases.

We compare these metrics of rarity for \hp
to those reported by Roy et al.~\cite{bugsynth} for
41 manually seeded bugs in the TCAS benchmark~\cite{infrastructure},
as well as 82~synthetic bugs seeded using their Apocalypse system in the same TCAS programs.
More precisely, Table~4 in~\cite{bugsynth} reports the number of all bug-triggering tests
generated by \klee within 1 hour, which corresponds to measure $t$.
Figure~5 in ~\cite{bugsynth} plots the number of tests generated by \klee before hitting a first bug,
which corresponds to measure $f$.
We directly compare these to the same measures on \hp's PUTs, 
without repeating \cite{bugsynth}'s experiments.
We only use \klee to investigate rarity
both because it is a standard choice for this kind of assessment~\cite{bugsynth},
and because its systematic exploration of program paths provides a more robust measure 
than others (such as testing time) that are strongly affected by the sheer size
and complexity of the PUT as a whole---as opposed to its bugs' specifically.

\subsection{RQ4: Capabilities}
\label{sec:design:capable}

To further demonstrate the flexibility of \hp's generation,
we look more closely at how different bug-finding tools
perform on different batches of PUTs generated by \hp.
Which PUTs are easier or harder to analyze
suggests which capabilities of the bug-finding tools
are more or less effective to analyze programs
with certain features.

We ran each of the tools \AFL, \CBMC, and \klee
on the PUTs in $B_{\tr{IC}}$, $B_{\tr{SC}}$, $B_{\tr{FL}}$, $B_{\tr{PC}}$, $B_{\tr{CC}}$, and $B_{\star}$.
Since these batches include multiple repetitions of the same transformation,
they demonstrate the generation of PUTs with homogeneous characteristics.
By observing how each tool's bug-finding capabilities
change in different batches, and within each batch as the same transformation is repeated multiple times,
we can outline each tool's strengths and weaknesses
in comparison with the other tools'
and link them to the characteristics of the transformations.

\section{Experimental Results}
\label{sec:exp-results}

\subsection{RQ1: Fairness}
\label{sec:results:fair}

\begin{table}[!tb]
  \centering
  \footnotesize
  \setlength{\tabcolsep}{4pt}
  \begin{tabular}{l *{7}{rr|} rr}
    \toprule
    \multicolumn{1}{c}{\textsc{batch}}
    & \textsc{\%} & \textsc{\#}
    & \textsc{\%} & \textsc{\#}
    & \textsc{\%} & \textsc{\#}
    & \textsc{\%} & \textsc{\#}
    & \textsc{\%} & \textsc{\#}
    & \textsc{\%} & \textsc{\#}
    & \textsc{\%} & \textsc{\#}
    & \textsc{\%} & \textsc{\#}
    \\
    \midrule
    $B_1$ & 0.0\% & 0 & 0.0\% & 0 & 0.0\% & 0 & 0.0\% & 0 & 0.0\% & 0 & 10.0\% & 1 & 90.0\% & 9 & 0.0\% & 0
    \\
    $B_2$ & 0.0\% & 0 & 2.2\% & 1 & 0.0\% & 0 & 15.6\% & 7 & 17.8\% & 8 & 0.0\% & 0 & 64.4\% & 29 & 0.0\% & 0
    \\
    $B_{10}$ & 22.0\% & 44 & 12.0\% & 24 & 6.5\% & 13 & 4.0\% & 8 & 8.5\% & 17 & 0.5\% & 1 & 5.5\% & 11 & 41.0\% & 82
    \\
    $B_{100}$ & 0.0\% & 0 & 0.0\% & 0 & 0.0\% & 0 & 0.0\% & 0 & 0.0\% & 0 & 0.0\% & 0 & 0.0\% & 0 & 100.0\% & 100
    \\
    $B_{1000}$ & 0.0\% & 0 & 0.0\% & 0 & 0.0\% & 0 & 0.0\% & 0 & 0.0\% & 0 & 0.0\% & 0 & 0.0\% & 0 & 100.0\% & 100
    \\
    \midrule
    $B$ & 9.7\% & 44 & 5.5\% & 25 & 2.9\% & 13 & 3.3\% & 15 & 5.4\% & 25 & 0.4\% & 2 & 10.8\% & 49 & 62.0\% & 282
    \\
    \midrule
     $A$ & \multicolumn{2}{c}{\setIn} & \multicolumn{2}{c}{\setOut} & \multicolumn{2}{c}{\setOut} & \multicolumn{2}{c}{\setIn} & \multicolumn{2}{c}{\setIn} & \multicolumn{2}{c}{\setOut} & \multicolumn{2}{c}{\setIn} &  \multicolumn{2}{c}{\setOut} 
    \\
     $C$ & \multicolumn{2}{c}{\setOut} & \multicolumn{2}{c}{\setIn} & \multicolumn{2}{c}{\setOut} & \multicolumn{2}{c}{\setIn} & \multicolumn{2}{c}{\setOut} & \multicolumn{2}{c}{\setIn} & \multicolumn{2}{c}{\setIn} & \multicolumn{2}{c}{\setOut}
    \\
     $K$ & \multicolumn{2}{c}{\setOut} & \multicolumn{2}{c}{\setOut} & \multicolumn{2}{c}{\setIn} & \multicolumn{2}{c}{\setOut} & \multicolumn{2}{c}{\setIn} & \multicolumn{2}{c}{\setIn} & \multicolumn{2}{c}{\setIn} & \multicolumn{2}{c}{\setOut}
    \\
    \bottomrule
  \end{tabular}
  \caption{For each combination of tools (those marked by \setIn{} in each column),
    for each \textsc{batch} of PUTs used in the experiments,
    the percentage \textsc{\%} and the absolute number \textsc{\#} of PUTs in the batch
    whose unique bugs were
    triggered exclusively by the tests generated by those tools.
    For example, the leftmost column indicates that tool $A$ managed to find bugs in 44 PUTs in batch $B$ (9.7\% of all PUTs in $B$), which no other tool could find.}
  \label{tab:fairness}
\end{table}

\autoref{tab:fairness} reports, for each batch of PUTs in \autoref{tab:batches},
which testing tools were able to
generate inputs triggering the PUTs' unique bugs
in our experiments.
Row $B$ corresponds to all PUTs used in these experiments.
At least one of the tools $A$, $C$, and $K$ managed to detect
bugs in 67.8\% of all PUTs with less than 100 transformations.%
\footnote{Corresponding to batch $B_{1} \cup B_{2} \cup B_{10}$.} %
The distribution is not strongly biased in favor of any tool%
---even though $A$ was noticeably more effective than $K$ and $C$,
as it was the only tool capable of detecting the bugs in 9.7\% of all PUTs.
On the other hand, every tool was somewhat effective,
and all three of them detected 10.8\% of the bugs.

Among the individual batches of PUTs,
$B_{10}$ is the ``fairest'',
in that it includes PUTs that are challenging for each individual testing tool.
In contrast, the PUTs in batches $B_1$ and $B_2$
are generally simple to analyze for most of the tools;
and the PUTs in batches $B_{100}$ and $B_{1000}$ are overly
complex, so much that no testing tool could detect their bugs in the allotted time.
These results are a consequence of the different parameters chosen to
create the PUTs in these batches. %
Overall,
these results suggests that \hp can generate PUTs with bugs that are \emph{fair},
as they are a mix of elusive (highly challenging) bugs
and simpler bugs that most practical testing frameworks can discover.

\subsection{RQ2: Reproducibility}
\label{sec:results:reproducible}

As expected, all PUTs produced by \hp for our experiments
passed the reproducibility checks discussed in \autoref{sec:design:reproducible}.
Namely:

\begin{enumerate}
\item Running each PUT on \hp's generated input triggers the unique bug in the PUT.
\item The PUTs compile without errors or warnings.
\item The PUTs behave in the same way regardless of which compiler is used to compile them.
\item LLVM's Undefined Behavior Sanitizer does not report any source of undefined behavior
  in the PUTs.
\end{enumerate}

These checks confirm that \hp produces PUTs with reproducible
seeded bugs, since they are well-formed and behave consistently as expected.

\subsection{RQ3: Depth and Rarity}
\label{sec:results:deep-rare}

\begin{figure}[!b]
  \centering
  \begin{subfigure}[b]{0.49\textwidth}
    \centering
    \includegraphics[width=1.0\linewidth]{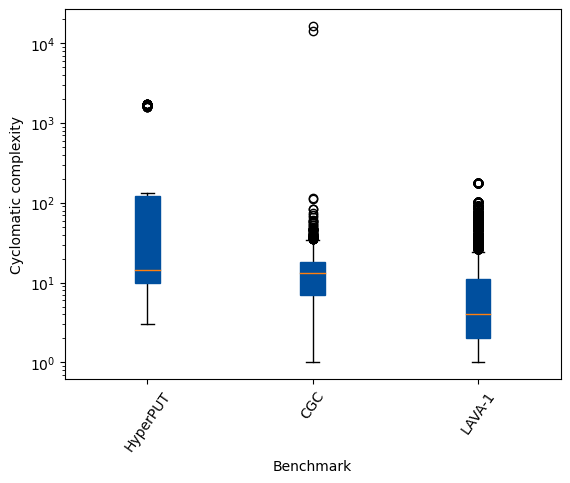}
    \caption{Box plots of the distributions of cyclomatic complexity per function.
      The vertical axis uses a logarithmic scale.}
    \label{fig:cyclomatic:boxplot}
  \end{subfigure}
  \hfill
  \begin{subfigure}[b]{0.44\textwidth}
    \centering
    \footnotesize
    	\begin{tabularx}{1.0\linewidth}{lrrr}
        \toprule
         & \textsc{\hp} & \textsc{cgc} & \textsc{lava-1} \\
        \midrule
        Mean & 444 & 14 & 12 \\
        Median & 18 & 13 & 4 \\
        Stddev & 727 & 144 & 20 \\
        Min & 3 & 1 &  1 \\
        Max & 1902 & 16386 & 179 \\
        \midrule
        Functions  & 455 & 22893 & 18906 \\
        \bottomrule
      \end{tabularx}
    \caption{Statistics about the distributions of cyclomatic complexity per function.}
    \label{tab:cyclomatic:stats}
 \end{subfigure}
 \caption{Distributions of cyclomatic complexity per function in three collections of buggy programs:
      the PUTs in batch $B$ generated by \hp, and benchmarks CGC~\cite{cgc} and LAVA-1~\cite{lava}.}
  \label{fig:cyclomatic}
\end{figure}

\begin{figure}[!bt]
  \centering
  \begin{subfigure}{0.49\textwidth}
    \centering
    \includegraphics[width=1.0\linewidth]{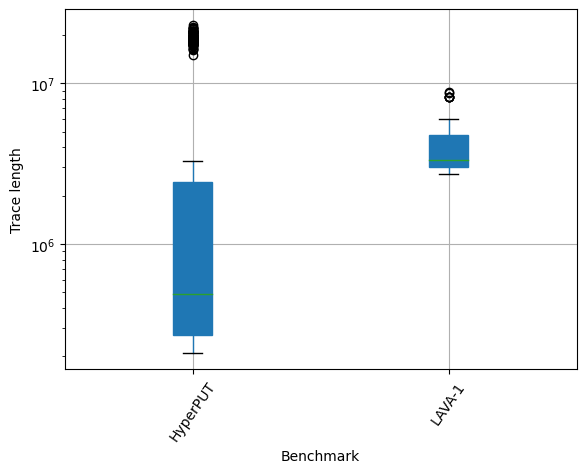}
    \caption{Box plots of the distributions of path length per bug. The vertical axis uses a logarithmic scale.}
    \label{fig:path-length:boxplot}
  \end{subfigure}
  \hfill
  \begin{subfigure}{0.44\textwidth}
    \setlength{\tabcolsep}{8pt}
    \small
    	\begin{tabular}{lrr}
		\toprule
		 & \textsc{\hp} & \textsc{lava-1} \\
        \midrule
        Mean & \numprint{4858702} & \numprint{4108228} \\
        Median & \numprint{486500} & \numprint{3339297} \\
        Stddev & \numprint{7710406} & \numprint{1644766} \\
        Min & \numprint{210260} & \numprint{2728637} \\
        Max & \numprint{22936332} & \numprint{8775398} \\
        \midrule
        Bugs  & 455 & 69 \\
        \bottomrule
      \end{tabular}
    \caption{Statistics about the distributions of path length per injected bug.}
    \label{tab:path-length:stats}
 \end{subfigure}
 \caption{Distributions of the length of the execution path on a bug-triggering input
   in two collections of buggy programs:
   the PUTs in batch $B$ generated by \hp, and benchmark LAVA-1~\cite{lava}.}
  \label{fig:path-length}
\end{figure}

\subsubsection{Depth}
\label{sec:results:deep}

\autoref{fig:cyclomatic}
summarizes the distribution of \emph{cyclomatic complexity} measures
for the functions in the PUTs generated by \hp (batch $B$),
and compares it to the functions featuring in the benchmarks CGC and LAVA-1.
\hp can generate very complex PUTs according to this metric:
even though some of CGC's programs are an order of magnitude more complex,
\hp's PUTs cover a broad range of cyclomatic complexities,
and are those with the highest average complexity.
This is a consequence of the way we configured \hp to generate
also large and complex PUTs in batches $B_{100}$ and $B_{1000}$ (as described in \autoref{sec:subjects}).
It suggests that \hp is capable of generating simple as well as complex PUTs,
and hence can generate a diverse collection of synthetic buggy programs.

Cyclomatic complexity measures the branching complexity of
programs, which is only a proxy for the complexity of
the \emph{bugs} that appear in the programs.
In principle, a very complex program may have very shallow bugs
if they occur in the first few lines of executable code.
Path length%
---the number of instructions executed from program entry until the bug is triggered---%
better assesses the depth of the synthetic bugs in \hp's generated PUTs.
\autoref{fig:path-length}
summarizes the distribution of \emph{path length} for each bug in the PUTs generated by \hp (batch $B$),
and compares it to the path length of synthetic bugs in the benchmark LAVA-1.

\hp's synthetic bugs are deeper on average (mean),
but LAVA-1's bugs are not that far behind, and have a much higher median.
In fact, \hp's have a higher standard deviation,
as the batch $B$ includes both small PUTs with
shallow short-path bugs and large PUTs with bugs that are deeply nested.

As for other measures,
this variety is a direct consequence of the way we configured \hp (as described in \autoref{sec:subjects}).
Overall, \hp can generate shallow as well as deep bugs,
including several that exhibit metrics similar to those of organic bugs.

\begin{table}[!bt]
  \centering
  \begin{subtable}{0.47\textwidth}
    \footnotesize
    	\begin{tabularx}{1.0\linewidth}{lrrr}
        \toprule
        & & \multicolumn{2}{c}{\cite[Fig.~5]{bugsynth}}
            \\
		& \textsc{\hp} & \textsc{tcas} & \textsc{Apocalypse} \\
        \midrule
        Mean & \numprint{7888} & \numprint{23} & \numprint{345} \\
        Median & \numprint{3575} & \numprint{17} & \numprint{165} \\
        Stddev & \numprint{12240} & \numprint{22} & \numprint{569} \\
        Min & \numprint{29} & \numprint{8} & \numprint{7} \\
        Max & \numprint{53389} & \numprint{152} &  \numprint{4366} \\
        \midrule
        Bugs & \numprint{20} & \numprint{41} &  \numprint{82} \\
		\bottomrule
	\end{tabularx}
   \caption{Statistics about the number $f$ of all test inputs generated by \klee per bug before triggering the bug in:
     \hp's batch $B_{\geq 6}$, manually seeded bugs in \textsc{TCAS}, and synthetic bugs seeded with \textsc{Apocalypse};
     the latter two are after \cite[Fig.~5]{bugsynth}.}
    \label{tab:rarity:tests-until-crash}
 \end{subtable}
  \hfill
  \begin{subtable}{0.49\textwidth}
    \footnotesize
    	\begin{tabularx}{1.0\linewidth}{lrrr}
        \toprule
        & & \multicolumn{2}{c}{\cite[Tab.~4]{bugsynth}}
            \\
		& \textsc{\hp} & \textsc{tcas} & \textsc{Apocalypse} \\
        \midrule
        Mean & \numprint{7655} (\numprint{409}) & 363 & 13 \\
        Median & \numprint{2518} (\numprint{64}) & 213 & 1 \\
        Stddev & \numprint{10782} (\numprint{541}) & 431 & 51  \\
        Min & \numprint{1} (\numprint{1}) & 24 & 1 \\
        Max & \numprint{40708} (\numprint{1402}) & 1805 &  341 \\
        \midrule
        Bugs & \numprint{20} & \numprint{41} &  \numprint{82} \\
		\bottomrule
	\end{tabularx}
   \caption{Statistics about the number $t$ of bug-trig\-ger\-ing test inputs per bug generated by \klee:
     \hp's batch $B_{\geq 6}$, manually seeded bugs in \textsc{TCAS}, and synthetic bugs seeded with \textsc{Apocalypse};
     the latter two are after \cite[Tab.~4]{bugsynth}.}
    \label{tab:rarity:triggering-tests}
 \end{subtable}
 \caption{Number of KLEE-generated inputs as a measure of bug rarity.}
  \label{tab:rarity}
\end{table}

\subsubsection{Rarity}
\label{sec:results:rare}

\autoref{tab:rarity} shows statistics about the rarity of bugs in \hp's PUTs
in $B_{\geq 6}$, and compares them to the analogous measures reported in Figure~5 and Table~4 of
Roy et al.~\cite{bugsynth}
about:
\begin{enumerate*}[label=\emph{\roman*})]
\item bugs in the TCAS benchmark, which consist of manually seeded bugs in several
  variants of an organic program;
\item bugs seeded using the Apocalypse system (introduced in~\cite{bugsynth})
  in the same programs of the TCAS benchmark.
\end{enumerate*}

\begin{figure}[!tbhp]
  \centering
  \begin{subfigure}{0.42\textwidth}
    \centering
    \includegraphics[width=1.0\linewidth]{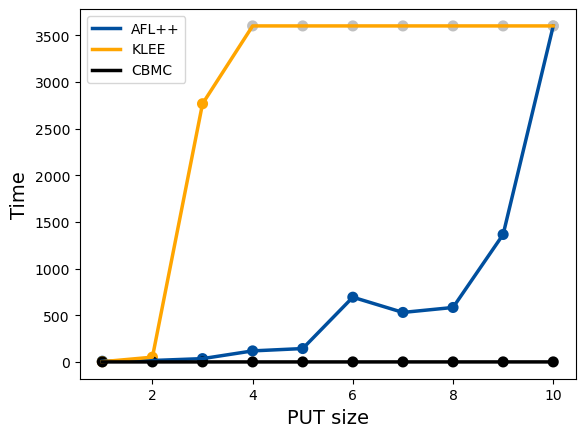}
    \caption{PUTs in batch $B_{\tr{IC}}$.}
    \label{fig:tr-IC}
  \end{subfigure}
  \hfill
  \begin{subfigure}{0.42\textwidth}
    \centering
    \includegraphics[width=1.0\linewidth]{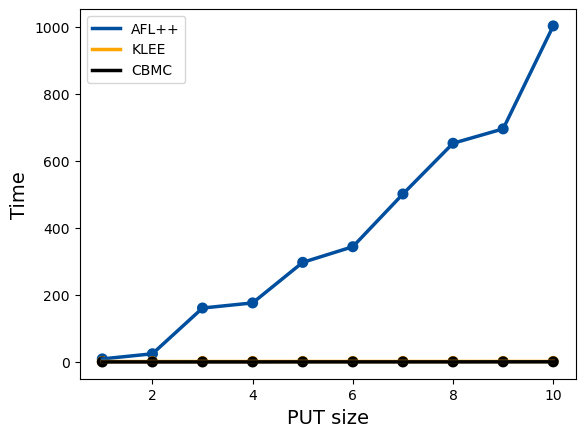}
    \caption{PUTs in batch $B_{\tr{SC}}$.}
    \label{fig:tr-SC}
  \end{subfigure}
  \\
    \begin{subfigure}{0.42\textwidth}
    \centering
    \includegraphics[width=1.0\linewidth]{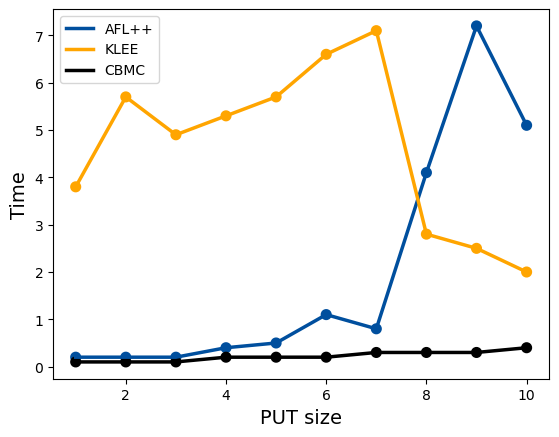}
    \caption{PUTs in batch $B_{\tr{FL}}$.}
    \label{fig:tr-FL}
  \end{subfigure}
  \hfill
  \begin{subfigure}{0.42\textwidth}
    \centering
    \includegraphics[width=1.0\linewidth]{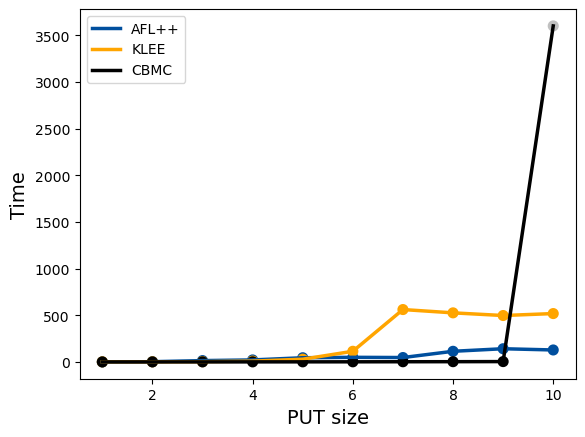}
    \caption{PUTs in batch $B_{\tr{PC}}$.}
    \label{fig:tr-PC}
  \end{subfigure}
  \\
  \begin{subfigure}{0.42\textwidth}
    \centering
    \includegraphics[width=1.0\linewidth]{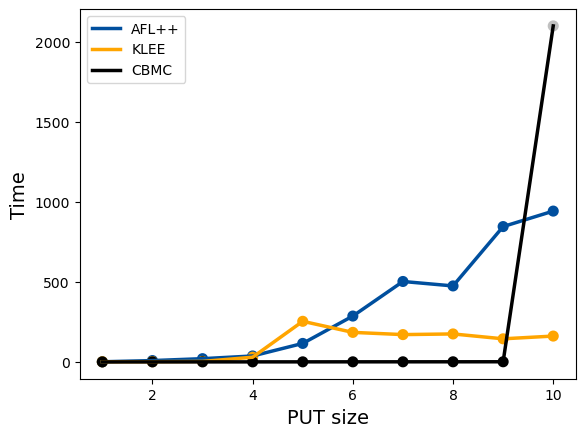}
    \caption{PUTs in batch $B_{\tr{CC}}$.}
    \label{fig:tr-CC}
  \end{subfigure}
  \hfill
  \begin{subfigure}{0.42\textwidth}
    \centering
    \includegraphics[width=1.0\linewidth]{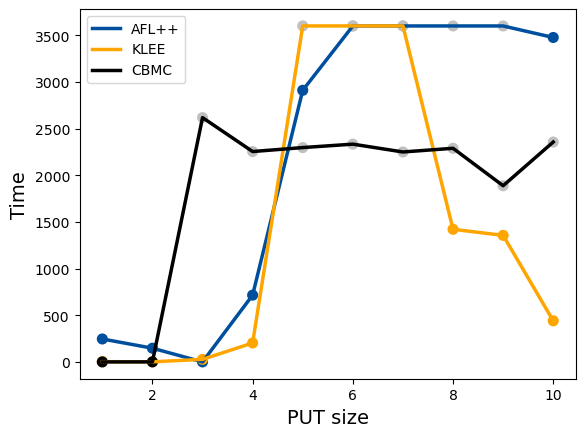}
    \caption{PUTs in batch $B_{\star}$.}
    \label{fig:tr-star}
  \end{subfigure}
  \caption{Running time to discover the bug in each PUT in batches $B_{\tr{IC}}, B_{\tr{SC}}, B_{\tr{FL}}, B_{\tr{PC}}, B_{\tr{CC}}, B_{\star}$. The horizontal axis enumerates the 10 PUTs in each batch in order of size (number of transformations). The vertical axis measures the running time (in seconds) until the tool terminates or times out (as in all other experiments, we report the average of 4 repeated runs).
    A colored filled disc indicates that the tool terminated successfully (it discovered the bug);
    a grayed out circle indicates that the tool terminated or timed out without discovering the bug.
    Data about \AFL are in color blue, about \CBMC are in color black, about \klee are in color yellow.
  }
  \label{fig:capable}
\end{figure}

As it can be seen, the average number of test cases is significantly higher for \hp. Consequently, the corresponding PUTs require considerably more queries before being correctly analyzed by the testing framework.

In principle, our generator also allows to always inject bugs with exactly a single triggering input, but for the purpose of achieving more variety in the process, several additional transformations with small parameters have been included. In addition, some transformations (such as the ones of type FL) allow the testing framework to easily generate new triggering inputs by randomly modifying the corresponding command-line argument $argv[i]$. For this reason the values in parenthesis were introduced. They refer to the number of triggering test cases for PUTs in Batch $B_{\geq 6}$ with non-negligible parameter size ($n>3$ for transformation $PC$ and $e>=90$ for transformation FL).

\subsection{RQ4: Capabilities}
\label{sec:results:capable}

The previous research questions
demonstrated that \hp is capable of producing PUTs
with bugs with a broad range of characteristics,
some comparable to those present in
commonly used benchmarks for bug-finding tools.
In particular, \autoref{sec:results:fair} suggests that
different PUTs are more or less challenging for different bug-finding tools.
In this section, we demonstrate how the variety of PUTs
generated by \hp can be used to exercise different
capabilities of bug-finding tools. %

To this end, we generated new batches of PUTs
$B_{\tr{IC}}, B_{\tr{SC}}, B_{\tr{FL}}, B_{\tr{PC}}, B_{\tr{CC}}, B_{\star}$.
As described in \autoref{sec:design:capable},
PUTs in each batch $B_T$ only use the same transformation $T$,
and differ only in their size---measured as the number of repetitions of $T$.
This way, we can understand how the characteristics of each transformation challenge a tool's
bug-finding capabilities.
\autoref{fig:capable} plots the running time of the considered testing frameworks when searching for bugs in these PUTs.
Unsurprisingly, the performance of a tool clearly depends on the transformations
that make up a PUT.
Let's look into each tool's performance on the different batches.

\CBMC is very effective on PUTs using transformations
\tr{IC}, \tr{SC}, and \tr{FL},
where it scales effortlessly.
PUTs using transformations \tr{IC} and \tr{SC}
have no loops, and hence \CBMC can easily build an exhaustive finite-state abstraction.
PUTs using transformations \tr{FL} do have loops, but in this case \CBMC manages to find a suitable loop unrolling bound
that makes the analysis exhaustive without blowing up the search space.
In contrast, 
\CBMC's performance suddenly blows up for the largest PUTs
using 10 transformations \tr{PC} and \tr{CC};
in these case, loops whose exit condition depends on 
an input string
become hard to summarize with a fixed, small unrolling bound past a certain size.
Similarly, \CBMC's performance on batch $B_\star$ depends on how many and %
which transformations are used;
in particular,
as soon as the randomly generated PUTs include several nested loops
with transformations \tr{PC} or \tr{CC}, \CBMC
runs out of resources and terminates in about 40 minutes without detecting the bugs.

\klee is as effective as \CBMC on PUTs using transformation \tr{SC}.
It outperforms \CBMC on PUTs using transformations \tr{PC} and \tr{CC},
where it scales graciously to the largest PUTs
thanks to its symbolic reasoning capabilities.
On PUTs using transformation \tr{FL}, \klee is always effective,
but its running times fluctuate somewhat unpredictably%
---albeit remaining reasonably low in absolute value.
This is probably a result of running \klee with randomized search
(see \autoref{tab:tf-configs}), a feature that can speed up the search for bugs
but also introduces random fluctuations from run to run.
In contrast, \klee struggles to scale on PUTs using transformation \tr{IC}
(both in batch $B_{\tr{IC}}$ and in batch $B_\star$).
The problem here is not the transformation per se,
but rather how it is instantiated in the PUTs generated for the experiments.
As we explain in \autoref{sec:subjects},
parameter $v_1$ in transformation \tr{IC}
is instantiated with \code{atoll(argv[i])},
which interprets a string command-line argument
as an integer;
since \klee does not have access to the source code of library function \code{atoll},
it treats it as a black box, and hence its constraint solving capabilities
are of little use to find efficiently a suitable string argument
that \code{atoll} converts to the integer $v_2$
(the transformation's second parameter, instantiated with a random integer).
This also explains the difference in performance with transformation \tr{SC},
where there is no black-box function involved, and hence \klee can
easily find a suitable input string from the transformation's condition itself.

\AFL remains reasonably effective largely independent of which transformations
are used; 
however, its running time tends to grow with the size of the analyzed PUT.
This behavior---complementary to \klee's and \CBMC's---is a result
of \AFL being a gray box tool.
In a nutshell,
this means that
\AFL does not have direct access to the source code of the analyzed functions;
thus, it cannot extract path constraints from it but has to ``guess'' them indirectly
by trial and error.
\AFL's gray-box strategy, combined with its many heuristics and optimizations,
achieves a different trade off than white-box tools like \klee and \CBMC:
\AFL is an overall more flexible tool (in that it is less dependent on the characteristics of the analyzed software), but usually requires more time
and has more random fluctuations in its behavior.
Another difference is in scalability:
\AFL's analysis time necessarily grows with the size of the inputs;
in contrast, symbolic techniques like \klee are much more insensitive
to input size, as long as the complexity of the symbolic constraints does not vary.

Overall, these results demonstrate how \hp can be used to
generate PUTs with heterogeneous characteristics and sizes,
which challenge different capabilities of
diverse bug-finding techniques.

\subsection{Limitations and Threats to Validity}
\label{sec:threats}

We discuss the main limitations of \hp's technique,
its current implementation, and other threats to
the validity of the experiments described in this section,
as well as how we mitigated them.

\paragraph{Construct validity}
depends on whether  
the measurements taken in the experiments reflect
the features that are being evaluated.
In our experiments, we mainly collected standard
measures, such as running time, whether a bug-finding tool
managed to trigger a bug, and static (cyclomatic complexity)
and dynamic (path length) measures of complexity.
For the experiments to answer RQ3,
we also counted the number of triggering test cases and generated test cases
for each bug---the same measures used by Roy et al.~\cite{bugsynth}
to assess bug rarity.
Using standard measures reduces the risk of threats to construct validity,
and helps ensure that our results are meaningfully comparable with
those in related work.

Our experiments to answer RQ4 were limited by the
transformations currently supported by \hp,
and by how we combined them.
These restrictions are still consistent with RQ4's aim,
which is to explore \hp's capabilities to exercise different testing techniques
with PUTs of different characteristics.

\paragraph{Internal validity}
depends on whether
the experiments adequately control for possible confounding factors.
One obvious threat follows from possible bugs in our implementation of \hp.
As usual, we mitigated this threat with standard software development practices,
such as (manual) regression testing, code reviews, and periodic revisions and refactoring.

To account for fluctuations
due to the nondeterministic/randomized behavior
of some testing tools,
we followed standard practices by
repeating each experiment multiple times,
and reporting the average values (see \autoref{sec:setup}).
We usually observed only a limited variance in the experiments,
which indicates that the practical impact of randomness was usually limited.

Our experiments ran with a timeout of one hour per analyzed bug;
it is possible that some experiments would
have resulted in success if they had been allowed a longer running time.
We chose this timeout as it is standard
in such experiments~\cite{bugsynth},
and compatible with running a good number of meaningful experiments in a reasonable time.
Our experiments
showed a considerable variety of behavior,
which suggests
that the testing tools we used can be successful within this timeout.

A related threat is in how we configured the testing tools
(see \autoref{tab:tf-configs}).
\AFL, \CBMC, and \klee
are highly-configurable tools, and their performance
can vary greatly depending on which options are selected.
Our goal was not an exhaustive exploration
of all capabilities of these tools, but rather
a demonstration of their ``average'' behavior.
Correspondingly, we mitigated this threat by:
\begin{enumerate*}[label=\emph{\roman*})]
\item running each tool with two configurations;
\item including the default configuration (with no overriding of default options);
\item using common, widely used options.
\end{enumerate*}

To answer RQ3 in \autoref{sec:design:rare},
we compared some measures taken on PUTs generated by \hp
with the same measures reported by Roy et al.~\cite{bugsynth}.
Since we did not repeat \cite{bugsynth}'s experiments
in the same environment where we ran \hp,
we cannot make strong, quantitative claims about the results of this comparison.
This limitation does not, however, significantly threaten
our overall answer to RQ3,
which is that \hp can generate bugs
whose rarity is realistic.
\cite{bugsynth}'s experiments
are used as a reference for what ``realistic'' means,
whereas our work's aims are largely complementary.

\paragraph{External validity}
depends on whether the experimental results generalize, and to what extent.

\hp currently generates PUTs with a trivial modular structure,
consisting of a single function that only uses a handful of
standard C libraries.
On the other hand, each function can be structurally quite intricate,
with bugs nested deep in the function's control-flow structure.
This is partly a limitation of the current implementation,
but also an attempt to focus on generating PUTs that are
\emph{complementary} to organic bug-seeded programs.
Detecting ``deep'' bugs is a relevant open challenge in test automation~\cite{surveyimp},
and synthetic buggy programs may be interesting subjects
to demonstrate progress in addressing the challenge.

\hp generates programs in C since this
is a widely popular target for the research on automated
testing and fuzzing.
The ideas behind \hp can certainly be applied to other programming languages,
possibly with different results.

Similarly, the choice of transformations currently supported by \hp
obviously limits its broader applicability.
\hp's implementation is extensible with new transformations;
deciding which ones to add depends on the goal
of the experiments one would like to  make.

\section{Conclusions}
\label{sec:discussion}

In this paper, we presented \hp, a technique and tool
to generate PUTs (Program Under Tests) with seeded bugs
automatically, according to desired characteristics.
The PUTs generated by \hp can be useful
as experimental subjects to assess the capabilities
of bug-finding tools, and how they change according to the
characteristics of the analyzed PUT.
To demonstrate this, we generated hundreds of PUTs
using \hp, and ran the popular bug-finding tools
\AFL, \CBMC, and \klee on them.
Our experiments suggest that \hp can generate heterogeneous
collections of PUTs, with several characteristics
that resemble those of ``ecologically valid'' bugs~\cite{bugsynth}.

The implementation of \hp is extensible,
so that users can easily add transformations
and parameters to configure the generation of bugs according to
the intended usage.
As future work, we plan to further
extend the flexibility of \hp,
so that it can also generate programs consisting of multiple functions and files,
or it can extend an existing program with new functions and seeded bugs.

\section*{Declaration of competing interest}
The authors declare that they have no known competing financial interests or personal relationships that could have appeared to influence the work reported in this paper.

\section*{Acknowledgments}
The authors gratefully acknowledge the financial support of the Swiss National Science Foundation for the project (SNF Grant Number 200020-188613).

\section*{Version of Record}
This preprint has not undergone peer review or any post-submission improvements or corrections. The Version of Record of this article is published in Empirical Software Engineering, and is available online at: \url{https://doi.org/10.1007/s10664-023-10430-8}.

%%%

%\bibliographystyle{plain}
%\bibliography{put_generator2}

\end{document}